\documentclass[prl,showpacs,amsmath,amssymb,twocolumn, 10pt]{revtex4}
\usepackage{enumerate}
\usepackage{color}
\usepackage{amsthm}
\usepackage{dcolumn}
\usepackage{bm}
\usepackage{graphicx}

\begin{document}

\newtheorem{corollary}{Corollary}
\newtheorem{definition}{Definition}
\newtheorem{example}{Example}
\newtheorem{lemma}{Lemma}
\newtheorem{proposition}{Proposition}
\newtheorem{theorem}{Theorem}
\newtheorem{fact}{Fact}
\newtheorem{property}{Property}
\newcommand{\bra}[1]{\langle #1|}
\newcommand{\ket}[1]{|#1\rangle}
\newcommand{\braket}[3]{\langle #1|#2|#3\rangle}
%%inner product
\newcommand{\ip}[2]{\langle #1|#2\rangle}
%%outer product
\newcommand{\op}[2]{|#1\rangle \langle #2|}

\newcommand{\tr}{{\rm tr}}
\newcommand{\supp}{{\it supp}}
\newcommand{\sch}{{\it Sch}}

\newcommand{\Span}{\mathrm{span}}
\newcommand {\E } {{\mathcal{E}}}
\newcommand{\In}{\mathrm{in}}
\newcommand{\Out}{\mathrm{out}}
\newcommand{\local}{\mathrm{local}}
\newcommand {\F } {{\mathcal{F}}}
\newcommand {\diag } {{\rm diag}}
\renewcommand{\b}{\mathcal{B}}
\newcommand{\h}{\mathcal{H}}
\renewcommand{\Re}{\mathrm{Re}}
\renewcommand{\Im}{\mathrm{Im}}
\newcommand{\Z}{\sigma_z}
\newcommand{\Clocal}{C^{(0)}_{\local}}
\newcommand{\Sp}[1][p]{S_{\min}^{(#1)}}

\title{Entanglement between Two Uses of a Noisy Multipartite Quantum
Channel Enables Perfect Transmission of Classical Information}
\author{Runyao Duan$^{1,2}$}
\email{dry@tsinghua.edu.cn}
\author{Yaoyun Shi$^{2}$}
\email{shiyy@eecs.umich.edu} \affiliation{$^1$State Key Laboratory
of Intelligent Technology and Systems,Tsinghua National Laboratory
for Information Science and Technology, Department of Computer
Science and Technology, Tsinghua University, Beijing 100084, China
\\
$^2$Department of Electrical Engineering and Computer Science,
University of Michigan, 2260 Hayward Street, Ann Arbor, MI
48109-2121, USA}
\date{\today}

\begin{abstract}
Suppose that $m$ senders want to transmit classical information to
$n$ receivers with zero probability of error using a noisy
multipartite communication channel. The senders are allowed to
exchange classical, but not quantum, messages among themselves, and
the same holds for the receivers. If the channel is classical, a
single use can transmit information if and only if multiple uses
can. In sharp contrast, we exhibit, for each $m$ and $n$ with $m\ge
2$ or $n\ge 2$, a quantum channel of which a single use is not able
to transmit information yet two uses can. This latter property
requires and is enabled by quantum entanglement.
\end{abstract}


\maketitle

The maximum rate at which a communication channel
carries information is characterized by a quantity
called its {\em channel capacity}. Different constraints
on the channel give rise to variants of channel capacity.
For example, the most commonly studied capacity, referred to
simply as the Shannon capacity,
can be attained using encoding and decoding
on a large number of uses of the channel, with the error probability
approaching $0$.

In 1956 Shannon introduced the notion of zero-error capacity to
characterize the ability of noisy channels to transmit classical
information with zero probability of error \cite{SHA56}. The study
of this notion and the related topics has since then grown into a
vast field called zero-error information theory \cite{KO98}, partly
motivated by the fact that no error may be allowed in some
applications, or the channel is not available for the large number
of uses required for attaining a small error probability. Unlike
Shannon capacity, the calculation of zero-error capacity is in
essence a combinatorial optimization problem on graphs and may be
extremely difficult even for very simple graphs. This difficulty has
in turn stimulated a great deal of research in graph theory (e.g.
\cite{LOV79}), a recent highlight of which is the resolution of the
Strong Perfect Graph Conjecture~\cite{CRST06}.

Since the physical process underlying all communication channels,
such as optical fibers, is quantum mechanical, a quantum information
theory is necessary to capture the full potential of communication
channels. While much progress has been made on understanding the
quantum analogy of Shannon capacity, little is known about quantum
zero-error capacity for transmitting classical information. Some
basic facts on the latter subject were observed in Ref. \cite{MA05},
while it was shown that the capacity is in general also extremely
difficult to compute~\cite{BS07}.

The purpose of this Letter is to demonstrate that quantum zero-error
capacity behaves dramatically different from the corresponding
classical capacity, and the difference is due to the effect of
quantum entanglement. We consider the following ``multi-user''
scenario: a set of $m$ senders want to send classical information
with zero probability of error to a set of $n$ receivers through a
noisy channel $\E$. We impose the following LOCC (Local Operations
and Classical Communication) requirement: The senders are allowed to
exchange classical, but not quantum, messages, and the same holds
for the receivers. Note that if $\E$ is a classical channel, the
LOCC restriction does not reduce the capacity. However, when $\E$ is
a quantum channel, the LOCC requirement may reduce the capacity.
Multi-user channels in which no communication is allowed within
senders and receivers have been studied widely in classical
information theory (see, e.g., Chapter 14 of~Ref. \cite{CT91}). Our
definition of multi-user channels is a natural extension of such
channels, and captures realistic settings where quantum
communication is expensive. Multipartite quantum communication was
studied in Ref. \cite{DCH04}. The scenario considered there differs
from ours in several important aspects: in their model, (a) the
quantum channel may be assisted with one- or two-way classical
channels; (b) their purpose was to study the additivity of the
channel capacity for transmitting {\em quantum} information with
vanishing error probability, while our focus is on the capacity for
transmitting classical information with zero-error probability.

\begin{figure}[ht]
  \centering
  \includegraphics[scale=0.4]{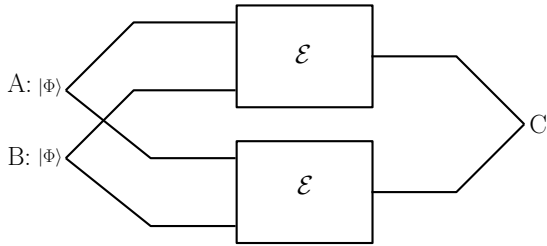}
  \caption{
  $\E$ is a noisy quantum channel from Alice and Bob to Charles. With separable input states,
  Alice and Bob cannot transmit classical information to Charles perfectly.
  Interestingly, by using $\E$ twice, the senders can transmit a classical bit perfectly to
  the receiver. To do so, each sender first prepares a locally maximally entangled
state $\ket{\Phi}=1/2(\ket{00}+\ket{11}+\ket{22}+\ket{33})$. If
Alice wants to send $``0"$ to Charles, they simply apply $\E$ twice.
Instead, if Alice wants to send $``1"$ to Charles, she first
performs $U=\op{0}{0}-\op{1}{1}+\op{2}{2}-\op{3}{3}$ on one particle
of her maximally entangled state and then the senders use $\E$ two
times. Bob can also send one single bit perfectly to Charles in a
similar way.}
  \label{pic1}
\end{figure}

We now describe our main result. When $\E$ is classical, it is
straightforward to see that its capacity is $0$ if and only if one
use of $\E$ cannot transmit information. In sharp contrast, we show
that this is not true for quantum channels in general. In
particular, we construct a quantum channel $\E$, for each pair of
$m$ and $n$, $m\ge2$ or $n\ge2$, that one use of $\E$ is not able to
transmit information yet two uses can. The later property can be
achieved in two different ways: (1) The senders apply $\E$ to create
a maximally entangled state between the receivers. The receivers
then distinguish the output states of the second use of $\E$ by
teleportation. (2) The senders locally prepare maximally entangled
states between the two uses of $\E$. The effect of the second case
cannot be observed under the assumptions of Refs. \cite{MA05,BS07}
where only product input states between two uses are allowed. Fig.
\ref{pic1} demonstrates our construction for $m=2$ and $n=1$.

Our construction in Case (2) uses the notion of completely entangled
subspace, which has been studied by several authors recently
\cite{PAR04, HLW06, DFJY07,CMW07,WS07,GW07}. More precisely, for any
$m>1$, we construct a partition of an $m$-partite state space into
two orthogonal subspaces, each of which contains no nonzero product
state. Such partitions were found in Refs. \cite{DY07,CHL+08}, and
can be used to construct counterexamples to the additivity of
minimum output $p$-R$\acute{\rm e}$nyi entropy for $p$ close to $0$
\cite{CHL+08}. Unfortunately, previous partitions are not sufficient
for our purpose.

We define some notions necessary for describing our constructions.
The set of $n$-bit binary strings is denoted by $\{0, 1\}^n$. Denote
the complement of $x\in\{0, 1\}^n$ by $\bar x$. We associate each
Hilbert space $\h$ a fixed orthonormal basis, referred to as the
computational basis and is usually denoted by $\{|i\rangle: 0\le
i\le \dim(\h)-1\}$ or $\{|x\rangle:x\in\{0, 1\}^n\}$ when
$\dim(\h)=2^n$. The operator space on $\mathcal{H}$ is denoted by
$\mathcal{B}(\mathcal{H})$. If $R\in\b(\h)$, denote by $R^T$ its
transpose with respect to the computational basis. Denote the
application of an operator $U\in\b(\h_k)$ on the $k$'th component of
a multipartite system $\h_1\otimes\cdots\otimes\h_n$ by $U^k$.

Let $m,n\ge1$ be positive integers. Let $\{A_1,\ldots,A_m\}$ be a
set of senders and $\{B_1,\ldots, B_n\}$ be a set of receivers.
Their state spaces are $\mathcal{H}_{A_k}$, $k=1,\ldots,m$, and
$\mathcal{H}_{B_\ell}$, $\ell=1,\ldots,n$, respectively. Denote by
$\mathcal{H}_{\In}=\otimes_{k=1}^m \mathcal{H}_{A_k}$ and
$\mathcal{H}_{\Out}=\otimes_{k=1}^n\mathcal{H}_{B_k}$. An $(m, n)$
{\em multi-user quantum channel} $\E$ is a completely positive
trace-preserving map from $\mathcal{B}(\mathcal{H}_{\In})$ to
$\mathcal{B}(\mathcal{H}_{\Out})$, and is used as follows. The
senders start with $|0\rangle\otimes\cdots\otimes|0\rangle$, and
encode a message $k$ into a state
$\rho_k\in\mathcal{B}(\mathcal{H}_{\In})$ through an LOCC protocol.
The receivers receive $\E(\rho_k)$, and decode the message $k$ by
LOCC.

Define $\alpha_{\local}(\E)$ to be the maximum integer $N$ with
which there exist a set of states
$\rho_1,\ldots,\rho_N\in\mathcal{B}(\mathcal{H}_{\In})$ such that:
(a) Each $\rho_k$ can be locally prepared by the senders, and (b)
$\E(\rho_1),\ldots,\E(\rho_N)$ can be perfectly distinguished by the
receivers using LOCC. It follows from the linearity of
superoperators that a set $\{\rho_k: k=1,\ldots,N\}$ achieving
$\alpha_{\local}(\E)$ can be assumed without loss of generality to
be orthogonal product pure states. Intuitively, one use of $\E$ can
be used to transmit $\log \alpha_{\local}(\E)$ bits of classical
information perfectly. When $\alpha_{\local}(\E)=1$ it is clear that
by a single use of $\E$ the senders cannot transmit any classical
information to the receivers perfectly.

The {\em local zero-error classical capacity} of $\E$, $\Clocal(\E)$,
is defined as follows:
\begin{equation}\label{c0}
C_{\local}^{(0)}(\E)=\sup_{k\geq 1}\frac{\log_2
\alpha_{\local}(\E^{\otimes k})}{k}.
\end{equation}

Suppose that $\E$ is classical (a so-called memoryless stationary
channel), that is, $\E=\sum_k \braket{k}{\cdot}{k}\rho_k$ for some
states $\rho_k$ diagonalized under the computational basis
$\{\ket{k}\}$. Then $\alpha_{\local}(\E)=1$ if and only if for all
pairs of $k$ and $\ell$, $\rho_k\rho_\ell\ne0$. Thus
$\alpha_{\local}(\E)=1$ if and only if $\alpha_{\local}(\E^{\otimes
k})=1$ for any $k$. Therefore, $C^{(0)}_{\local}(\E)=0$ if and only
if $\alpha_{\local}(\E)=1$. Our main theorem is:

\begin{theorem}\upshape\label{thm:main}
For any $m$ and $n$ with $m\ge2$ or $n\ge2$, there exists a
multi-user channel $\E$ from $m$ senders to $n$ receivers such that
$\alpha_{\local}(\E)=1$ and $C^{(0)}_{\local}(\E)>0$.
\end{theorem}

{\bf Proof.} For any positive integers $m, n, m', n'$, an $(m, n)$
channel $\E$ can be extended to an $(m+m', n+n')$ channel $\E'$,
which ignores the inputs from the additional $m'$ senders and
provides $|0\rangle$ to all the additional $n'$ receivers. Then for
any $k$, $\alpha_{\local}(\E^{\otimes k})=
\alpha_{\local}(\E'^{\otimes k})$. Thus we need only to prove the
theorem for $(m, n)=(1, 2)$ and $(2, 1)$.

Let $\rho_0=|\alpha\rangle\langle\alpha|$ and
$\rho_1=\frac{1}{3}(I-\rho_0)$, where $|\alpha\rangle=
\frac{1}{\sqrt{2}}(|00\rangle+|11\rangle)$. Consider the following
$(1, 2)$ channel $\E_{12}$, which is from $1$ qubit to $2$ qubits:
\[ \E_{12}= \langle0|\cdot|0\rangle\rho_0+\langle1|\cdot|1\rangle\rho_1.\]
The only output pairs that are orthogonal are $\rho_0$ and $\rho_1$.
However, they cannot be distinguished by an LOCC protocol. Thus
$\alpha_{\local}(\E_{12})=1$. On the other hand, when $\E_{12}$ is
used twice, the two receivers, Alice and Bob, can distinguish the
output states $\{\rho_0\otimes\rho_0=\E_{12}^{\otimes
2}(|00\rangle\langle00|),
\rho_0\otimes\rho_1=\E_{12}^{\otimes2}(|01\rangle\langle01|)\}$ by
the following LOCC protocol: Alice first teleports her second qubit
to Bob using the first qubit (which is maximally entangled with the
first qubit of Bob), then Bob applies a local measurement to
distinguish $\rho_0$ and $\rho_1$. Thus it follows from Eq.
(\ref{c0}) that $\Clocal(\E_{12})\ge 0.5$.

Now we turn to the construction of a $(2,1)$ channel. The input
space $\h_{\In}=\h_A\otimes\h_B$, and the output space is $\h_C$.
Each of $\h_A$ and $\h_B$ is a $4$ dimensional space and $\h_C$ is a
qubit. Let $S_0\subseteq\h_{\In}$ be the subspace spanned by the
following vectors:
\begin{eqnarray}\label{S0}
\ket{\psi_1}&=&\ket{00}-\ket{11},\nonumber\\
\ket{\psi_2}&=&\ket{22}-\ket{33},\nonumber\\
\ket{\psi_3}&=&\ket{20}-\ket{31},\nonumber\\
\ket{\psi_4}&=&\ket{02}+\ket{13},\nonumber\\
\ket{\psi_5}&=&\ket{30}-\ket{03},\nonumber\\
\ket{\psi_6}&=&\ket{10}-\sqrt{2}\ket{21}+\ket{32},\nonumber\\
\ket{\psi_7}&=&\ket{01}+\sqrt{2}\ket{12}+\ket{23},\nonumber\\
\ket{\psi_8}&=&\ket{10}-\ket{32}-\ket{01}+\ket{23},
\end{eqnarray}
and $S_1=S_0^\perp$. Let $P_\ell$ be the projector onto $S_\ell$,
$\ell=0,1$. The channel $\E_{21}:\b(\h_{\In})\to\b(\h_C)$ is defined
as:
\begin{equation}\label{eqn:channelE}
\E_{21}(\rho) =
\tr(P_0\rho)|0\rangle\langle0|+\tr(P_1\rho)|1\rangle\langle1|.
\end{equation}
Let $U=\op{0}{0}-\op{1}{1}+\op{2}{2}-\op{3}{3}$.  The projections
$P_0$ and $P_1$ have the following useful properties:
\begin{enumerate}[(i)]
\item $P_\ell=P_{\ell}^T$ and $P_\ell=U^i P_{\bar \ell}U^i$, for
any $\ell\in\{0, 1\}$ and $i=A,B$. As a consequence, $P_\ell^T
P_{\bar \ell}=P_\ell^T U^i P_{\ell}=0$. (Note that $U^\dagger=U$).
\item Neither $S_0$ nor $S_1$ contains a product state, i.e.
both $S_0$ and $S_1$ are completely entangled.
\end{enumerate}

Property (i) can be verified by inspection. We now prove Property
(ii). Let $\ket{\Psi}=(\sum_{k=0}^3a_k\ket{k})^A\otimes
(\sum_{\ell=0}^3 b_\ell\ket{\ell})^B$ be a product vector orthogonal
to $\ket{\psi_k}$, $k=1,\ldots, 7$. Then
\begin{eqnarray}\label{eqns}
a_0b_0-a_1b_1&=&0,\nonumber\\
a_2b_2- a_3b_3&=&0,\nonumber\\
a_2 b_0- a_3b_1&=&0,\nonumber\\
a_0b_2+a_1b_3&=&0,\nonumber\\
a_3b_0-a_0b_3&=&0,\nonumber\\
a_1b_0-\sqrt{2}a_2b_1+a_3b_2&=&0,\nonumber\\
a_0b_1+\sqrt{2}a_1b_2+a_2b_3&=&0.
\end{eqnarray}
Suppose that $a_0b_0\ne 0$. Assume without loss of generality that
$a_0=b_0=1$. Then
\begin{eqnarray}
a_1b_1=1,~a_2b_2=a_3b_3,~a_2=a_3b_1,\nonumber \\
b_2=-a_1b_3,~a_3=b_3,~b_1=-\sqrt{2}a_1b_2-a_2b_3.
\end{eqnarray}
By $a_1b_1=1$ and $a_2=a_3b_1$ we have $a_3=a_1a_2$. Substituting
$b_2=-a_1b_3$ and $a_3= b_3$ into $a_2 b_2= a_3b_3$ we have
$-a_1a_2a_3=a_3^2$. If $a_3=0$ then we have $b_2=b_3=0$, which
together with $b_1=-\sqrt{2}a_1b_2-a_2b_3$ implies $b_1=0$. However,
this is impossible as we have $a_1 b_1=1$. Thus $a_3\neq 0$. This
together with $-a_1a_2a_3=a_3^2$ implies $a_3=-a_1a_2$. However, we
have already shown that $a_3=a_1a_2$. Thus $a_3=-a_3=0$, again a
contradiction. Therefore $a_0b_0=0$. Note that if $a_kb_\ell=0$ and
for a nonzero constant $\lambda$,  $\lambda
a_kb_{\ell'}=a_{k'}b_{\ell}$, then
$a_{k}b_{\ell'}=a_{k'}b_{\ell}=0$. Applying this inference rule many
times, one concludes that all $a_kb_\ell=0$, $0\le k, \ell\le 3$ in
both $a_0=0$ and $b_0=0$ cases. Thus $\ket{\Psi}=0$, and $S_1$
contains no product state. By Property (i), this implies that $S_0$
does not contain a product state, either.

It follows from Property (ii) that $\E_{21}(\op{\psi}{\psi})$ is a mixed
state for any product state $\ket{\psi}$. Thus
$\alpha_{\local}(\E_{21})=1$. We now consider using $\E_{21}$ twice. Let
$\ket{\Phi}=1/2(\ket{00}+\ket{11}+\ket{22}+\ket{33})$.
Define $\ket{\Psi_0}$ and $\ket{\Psi_1}$ as follows:
$$\ket{\Psi_0}=\ket{\Phi}^{AA'}\otimes \ket{\Phi}^{BB'}~\textrm{and}~~\ket{\Psi_{1, i}}=U^i\ket{\Psi_0},$$ where
$i\in\{A,A',B,B'\}$. Note that for any operators $R_1$ and $R_2$,
$\langle\Psi_0|R_1^{AB}\otimes
R_2^{A'B'}|\Psi_0\rangle=\tr((R_1)^TR_2)$. Thus for any
$\ell\in\{0, 1\}$, applying Property (i), we have
\begin{eqnarray*}
\braket{\Psi_0}{P_{\bar \ell}^{AB}\otimes
P_\ell^{A'B'}}{\Psi_0}=&\tr(P_{\bar \ell}^T P_\ell)=0,
\end{eqnarray*}
and
\begin{eqnarray*}
\braket{\Psi_{1, i}}{P_\ell^{AB}\otimes P_\ell^{A'B'}}{\Psi_{1,
i}}=\tr(P_\ell^T U^i P_\ell U^i)=0.
\end{eqnarray*}
Thus $\E_{21}^{\otimes2}(\op{\Psi_0}{\Psi_0})$ and
$\E_{21}^{\otimes2}(\op{\Psi_{1, i}}{\Psi_{1, i}})$ are orthogonal. This can
also be verified by the following facts:
\begin{eqnarray}
\E_{21}^{\otimes2}(\op{\Psi_0}{\Psi_0})&=&({\op{00}{00}+\op{11}{11}})/2,\nonumber\\
\E_{21}^{\otimes2}(\op{\Psi_{i, 1}}{\Psi_{1,
i}})&=&(\op{01}{01}+\op{10}{10})/2.\nonumber
\end{eqnarray}
Therefore $\alpha_{\local}(\E_{21}^{\otimes 2})\ge 2$, and
$C^{(0)}_{\local}(\E_{21})\ge 0.5$. \hfill$\blacksquare$

The quantum channel $\E_{21}$ constructed above has the following
desirable property: when it is used twice, each sender is able to
transmit one classical bit without leaking any information about
this bit to the other sender. This is based on the fact that on
$U^A|\Psi_0\rangle$ and $U^B|\Psi_0\rangle$, the channel outputs the
same state orthogonal to the output on $|\Psi_0\rangle$. While
tensoring $\E_{21}$ with trivial channels does not preserve this
property, we are able to construct a family of channels $\E_{m1}$,
$m\ge3$, that have this property. Each $\E_{m1}$ is a channel from
$m$ qubits to $1$ qubit defined in analogy to $\E_{21}$ with the
following set of base vectors for $S_0$ ($S_1=S_0^\perp$):
$$\ket{\psi_{0}}=\ket{0}^{\otimes m}+\ket{1}^{\otimes m},\quad\textrm{and,}$$
$$\ket{\psi_x}=\ket{0}\ket{x}-\ket{1}\ket{\bar x},\ \ x\in\{0, 1\}^{m-1},\ x\neq 0^{m-1}.$$
The proof for $\alpha_{\local}(\E_{m1})=1$ and
$\Clocal(\E_{m1})\ge0.5$ is similar to that for $\E_{21}$, thus
we leave it to the interested reader.

We note that if the ``privacy'' property is not required,
the input dimension of $\E_{21}$ can be reduced to
$3\otimes 4$ with the following set of base vectors for $S_0$:
\begin{eqnarray}
\ket{\psi'_1}&=&\ket{10}-\ket{21},\nonumber\\
\ket{\psi'_2}&=&\ket{02}+\ket{13},\nonumber\\
\ket{\psi'_3}&=&\ket{20}-\ket{03},\nonumber\\
\ket{\psi'_4}&=&\ket{00}-\sqrt{2}\ket{11}+\ket{22},\nonumber\\
\ket{\psi'_5}&=&\ket{01}+\sqrt{2}\ket{12}+\ket{23},\nonumber\\
\ket{\psi'_6}&=&\ket{00}-\ket{22}-\ket{01}+\ket{23}.
\end{eqnarray}
Now only Bob (the party with the $4$ dimensional state space) can
transmit a private bit with two uses of the channel. This is simply
due to the fact that both Properties (i) and (ii) still hold with
the restriction that $i=B$ in Property (i). We omit the details of
the proof as it is similar to that for $\E_{21}$.

Our construction of $\E_{21}$ and the above variant has another
application on the additivity of the minimum output $p$-R$\acute{\rm
e}$nyi entropy $\Sp$. For $0\le p\le \infty$ and a quantum channel
$\E$,  $\Sp(\E)$ is defined as
\[\Sp(\E)=\min_{\rho} \frac{1}{1-p} \log(\tr(\E(\rho)^p)),\]
where $\rho$ is a density operator, and at $p=0,1,+\infty$, the
right hand side takes the limit. The {\em additivity problem} on
$\Sp$ asks if
\begin{equation}\label{eqn:additivity}
\Sp(\E_1\otimes \E_2)=\Sp(\E_1)+\Sp(\E_2), \quad\forall \E_1, \E_2.
\end{equation}
The case of $p=1$ is a central open problem in quantum information
theory~\cite{Shor04}. This motivates the study of the same question
for other values of $p$ by many authors. It has been shown that
$\Sp$ is not additive for any $p>1$ \cite{HW02}. Very recently
Ref.~\cite{CHL+08} showed that $\Sp$ is not additive for $p$ in a
neighborhood of $0$, by constructing a counterexample for $p=0$.
This is done by first constructing two completely entangled
subspaces $S$ and $S'$ of a bipartite space so that $S\otimes S'$ is
not completely entangled. Two channels $\E_1$ and $\E_2$ are then
defined based on $S$ and $S'$, respectively, through the
Choi-Jamio\l{}kowski isomorphism. It remains open if $\Sp[0]$ is
additive when $\E_1=\E_2$. We answer this question negatively:
setting both $S$ and $S'$ to be the $S_0$ in $\E_{21}$ (and its
variant) gives a channel $\E$ so that
$\Sp[0](\E^{\otimes2})<2\Sp[0](\E)$. We omit the details of the
argument as it is similar to that in Ref. \cite{CHL+08}.

In conclusion, we have shown that for a broad class of multi-user
quantum channels, of which the communication within the senders and
receivers is restricted to be LOCC, a single use of the channel
cannot be used to transmit classical information with zero
probability of error, while multiple uses can. The latter property
requires, and is a consequence of, quantum entanglement between
different uses, thus cannot be achieved by classical channels.

For all the channels we know, the LOCC restriction on encoding and
decoding is necessary. It remains an intriguing open problem if such
channels exist for a single sender and a single receiver. We do not
know the exact values of $\Clocal$ for the constructed channels,
despite the simplicity of their definitions. The difficulty in
computing them is not unexpected given that $\Clocal$
generalizes the classical concept of zero-error capacity, which can
be notoriously difficult to compute even for simple channels.
Developing methods for estimating $\Clocal$ is thus of great
theoretical interest, besides its obvious practical usefulness.

We are grateful to Andreas Winter for very useful discussions at
AQIS2007 and for sending us a copy of Ref. \cite{CHL+08} prior to
publication. This work was partially supported by the National
Science Foundation of the United States under Awards~0347078 and
0622033. R. Duan was also partially supported by the National
Natural Science Foundation of China (Grant Nos. 60702080, 60736011,
60503001, and 60621062) and the Hi-Tech Research and Development
Program of China (863 project) (Grant No. 2006AA01Z102).

\end{document}